\documentclass[useAMS,usenatbib,usegraphicx]{mn2e}

\title[Circumstellar discs in NGC\,6611]{Circumstellar discs around solar-mass stars in NGC\,6611}
\author[J.M.\,Oliveira et al.]{J.M. Oliveira$^{1}$\thanks{E-mail:
joana@astro.keele.ac.uk}, R.D. Jeffries$^{1}$, J.Th. van Loon$^{1}$, S.P Littlefair$^{2,3}$, T. Naylor$^{2}$\\
$^{1}$School of Chemistry and Physics, Keele University, Keele, Staffordshire ST5 5BG \\
$^{2}$School of Physics, University of Exeter, Stocker Road, Exeter EX4 4QL\\
$^{3}$Department of Physics and Astronomy, University of Sheffield, Sheffield S3 7RH}

\begin{document}

\date{Accepted . Received ; in original form }

\pagerange{\pageref{firstpage}--\pageref{lastpage}} \pubyear{2004}

\maketitle

\label{firstpage}

\begin{abstract}
We have performed $IZJHKL'$ observations in NGC\,6611, the young cluster that ionises the Eagle Nebula. We have discovered a rich pre-main sequence concentrated around the O-stars in the cluster. As measured by their L$'$-band excesses, at least 58\%\,$\pm$\,5\% of the pre-main sequence objects (0.45\,M$_{\odot}$\,$<$\,M\,$<$\,2\,M$_{\odot}$) have circumstellar discs. By comparing this disc frequency with frequencies determined for regions where the pre-main sequence stars are subject to less ionising radiation, we find no evidence that the harsher environment of NGC\,6611 (approximately an order of magnitude more ionising Lyman continuum radiation than the Trapezium cluster) significantly hastens the dissipation of circumstellar discs around solar-mass stars. 
\end{abstract}


\section{Introduction}

A significant fraction of the low-mass stars in our Galaxy may be born in environments where very massive stars are also present. The winds and ionising radiation from O-stars could be expected to change the properties of low-mass stellar populations in two ways: by inhibiting low-mass star formation \citep[e.g.,\,][]{silk77} and/or by hastening the dispersal of their circumstellar discs \citep*{johnstone98}. However, observations in several nearby OB-associations (e.g., the Trapezium cluster) reveal a wealth of low-mass stellar objects, born from the same stellar nursery as the more massive stars. 

Circumstellar discs are an essential ingredient of low-mass star formation, yet they seem to disappear quickly in nearby star forming regions \citep*{haisch01b}. The timescale for discs to dissipate is crucial in determining whether planets form and on which timescales \citep{brandner00}, and might control planetesimal migration \citep{nelson00}. It is not known which process dominates disc dispersal with likely culprits being viscous accretion, radiation pressure from the central star and from external sources, dust agglomeration and planetesimal formation. There is observational evidence that luminous O-stars affect circumstellar discs around low-mass stars (e.g., the proplyds in Orion, \citealt*{odell93}), supported by theoretical models of disc photo-evaporation by UV radiation \citep*{matsuyama03}. Judging from L-band disc excesses, low-mass stars in young clusters exhibit extremely high disc frequencies ($\ga$\,80\%, e.g.\ the Trapezium cluster, \citealt{lada00}) up to ages of $\sim$\,1.5\,Myr, which then decrease rapidly with age: at $\sim$\,3\,Myr, 50\% of discs have been dissipated, and the timescale for all cluster members to lose their discs is $\sim$\,6\,Myr \citep{haisch01b}. Recently, \citet{stolte04} found that in the starburst cluster NGC\,3603 disc dissipation occurs on shorter timescales. It is still not clear whether photo-evaporation is an important contributor to this decline.

We are investigating the low-mass star and disc population in NGC\,6611, a young and massive cluster \citep{walker61} that is currently ionising the Eagle Nebula. It contains a dozen O-stars (earliest spectral type O4, \citealt*{bosch99}), that produce $\sim$\,7 times more ionising (Lyman continuum) radiation than the Trapezium cluster \citep{hillenbrand93}. The visual extinction towards NGC\,6611 is known to be variable \citep{dewinter97}. Estimated distances range between 2\,$-$\,2.6\,kpc and the cluster age is about 2\,$-$\,3\,Myr \citep[e.g.,\,][]{hillenbrand93}. We have obtained $IZJHKL'$ photometry of NGC\,6611 in order to identify the pre-main sequence (PMS) population and determine the cluster disc frequency. Our aim is to establish empirically whether the massive stars significantly hasten disc dissipation around solar-mass stars in this cluster.

\section{Observations and Data reduction}

The complete details of the observing strategy and the reduction procedure for the data sets described here will be discussed in a forthcoming paper (Oliveira et al. in preparation); a summary is presented here.

\begin{figure}
\includegraphics[height=8.cm]{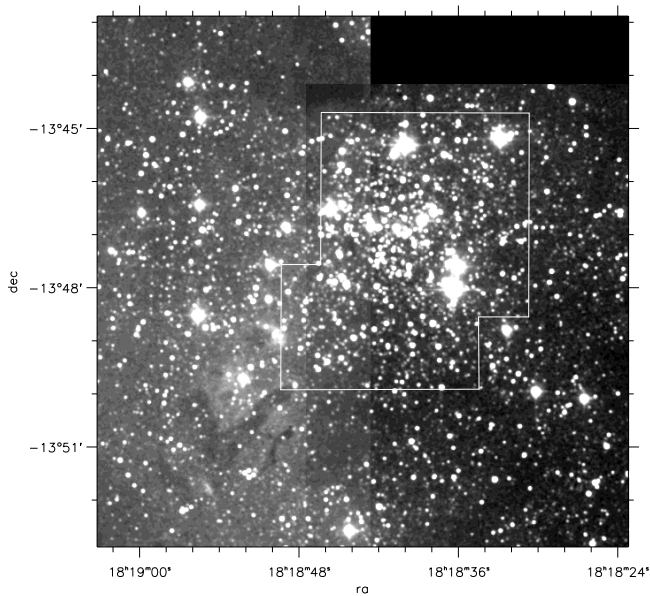}
\caption{A mosaic of 2MASS J-band images covering NGC\,6611 and the Eagle Nebula. Our $IZ$ survey covers the entire star forming region, while the $JHKL'$ survey covers the area delineated by the white lines.}
\label{image}
\end{figure} 

The region containing NGC\,6611 was surveyed using the WFC at the Isaac Newton Telescope. Observations were taken through Gunn i' and z' filters on the nights of 16$-$17 July 2002. The WFC consists of four CCD detectors; the camera has 0.33\,arcsec pixels and a field-of-view (FOV) of 34\,arcmin$\times$\,34\,arcmin (with an 11\,arcmin\,$\times$\,11\,arcmin square missing from the North-West corner). Total exposure times are 8\,$\times$\,300\,s and 8\,$\times$\,400\,s respectively at i' and z'; 20\,s frames were also obtained in each band to account for the high dynamic range of the objects' brightness. Landolt photometric standards were measured and these were used to obtain Cousins $I$ magnitudes and $(I-Z)$ colours. The data were reduced in a standard way (bias subtraction, flatfielding, de-fringing). The data analysis proceeded using the {\sc cluster} suite of software \citep{naylor02,jeffries04}. The external precision of the photometry was tested against $I$ magnitudes from the ESO EIS survey (S. Smartt and C. Evans, private communication); it reveals good systematic agreement to better than 0.05\,mag between 12\,$<$\,I\,$<$\,19\,mag (the faint limit arising from the EIS survey).

The $JHKL'$ photometry (MKO-IR system, \citealt*{tokunaga02}) of NGC\,6611 was obtained with UIST at the UK Infrared Telescope in June 2003 and June$-$August 2004. In Fig.\,\ref{image} we show the 6 fields that were observed at UKIRT, covering an approximate area of 4\,arcmin\,$\times$\,5\,arcmin (FOV 2\,arcmin\,$\times$\,2\,arcmin with 0.12\,arcsec pixels). The observations in the $JHK$ filters were performed in a standard 9-point dither pattern. The L$'$-band images required 55\,h of observations to cover the requested area. These observations were performed with a 4-point dither pattern and with chopping to a nearby field to properly sample the sky background. The L$'$-band images were taken in seeing better than 0.6\,arcsec and time on source varied between 60$-$90\,min, allowing us to reach $L'$\,$\sim$\,14\,mag with signal-to-noise of 10. The reduction was performed with {\sc orac-dr}, the UKIRT data reduction and high-level instrument control software. Multi-object photometry was performed on the $JHKL'$ images using an implementation of {\sc daophot} (version {\sc ii}) and {\sc allstar} \citep{stetson87} within ESO-MIDAS. The $JHK$ photometry was calibrated using the 2MASS catalogue while standard stars were observed to calibrate the L$'$-band photometry. The astrometry was also performed with relation to the 2MASS catalogue. 

\section{Pre-main sequence photometric candidates}

\begin{figure*}
\includegraphics[height=8.5cm]{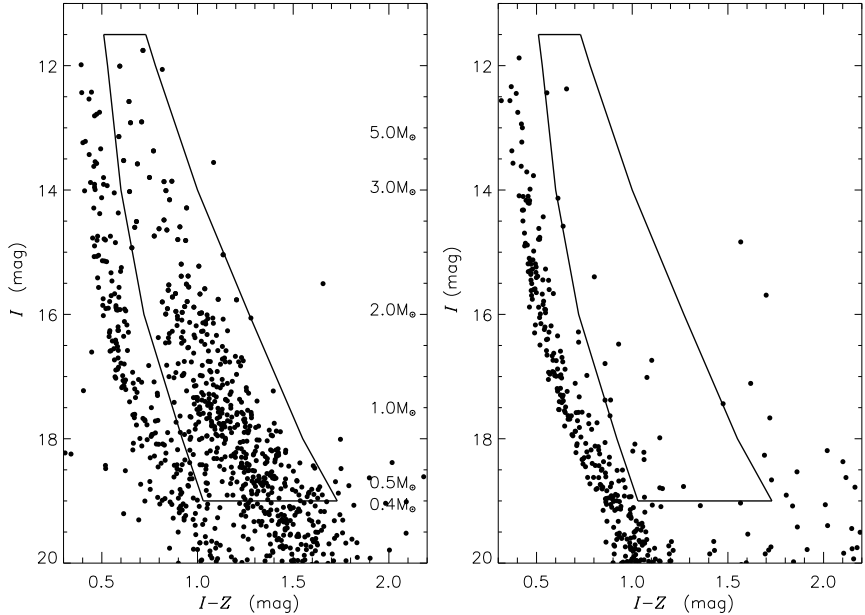}
\caption{$I$/($I-Z$) colour-magnitude diagram, for the central 7\,arcmin\,$\times$\,6\,arcmin area of NGC\,6611 (left) and for a control field 16\,arcmin West of NGC\,6611 (right). The PMS can be clearly seen in the NGC\,6611 field, separated from the bulk of field star contamination, but it is absent in the control field. The polygon indicates the selection area for a PMS object. Approximate PMS masses are indicated, using the models of \citet{siess00} and adopting a distance of 2\,kpc, $A_{\rm V}$\,$\sim$\,3.4\,mag and an age of 3\,Myr \citep{hillenbrand93}.}
\label{cmd}
\end{figure*}

The first step was to identify the pre-main-sequence population in NGC\,6611. Fig.\,\ref{cmd} shows the $I$/$(I-Z)$ colour-magnitude diagram for the central (7\,arcmin\,$\times$\,6\,arcmin) area of the cluster. PMS stars are redder than main-sequence stars of similar I-magnitude and therefore the PMS appears well separated from the bulk of foreground contamination in this diagram. The polygon represents the selection domain for a star to be considered a PMS candidate. Fig.\,\ref{cmd} also shows an equivalent diagram for an area of the same size situated 16\,arcmin West of the cluster. The PMS is effectively absent; we estimate foreground contamination to be of the order of 5\%. 

The fainter magnitude limit of the sample is imposed by the IR observations, i.e.\, we ensure that for the faintest stars in our sample ($I$\,=\,19\,mag) we are still able to detect stellar photospheric emission in the L$'$-band. At the lower mass end, the catalogue is essentially complete; however objects brighter than $I$\,$\sim$\,12\,mag are saturated in the I-band images. The uncertainties in magnitude and colour are generally much better than 0.1\,mag. We have identified 259 PMS candidates (with good $IZ$ photometry) in the area covered by our IR survey. We estimate that these objects have masses in the range 0.45\,M$_{\odot}$\,$\la$\,M\,$\la$\,7\,M$_{\odot}$ (using models by \citealt*{siess00}), assuming a distance of 2\,kpc, an average extinction $A_{\rm V}$\,$\sim$\,3.4\,mag and an age of 3\,Myr \citep{hillenbrand93}.

\section{Circumstellar discs}

Circumstellar dust discs are cooler than the stellar photospheres therefore 
they radiate mainly at IR wavelengths. If a young star has an IR colour in excess of the stellar photosphere this indicates the presence of a circumstellar disc. Traditionally, $(H-K)$ excesses were used as a disc indicator, but such an excess is only detectable for discs with high accretion rates and low inclinations \citep{hillenbrand98}. In the L$'$-band (or at longer wavelengths) the contrast between the stellar photosphere and disc emission is large even for relatively small disc masses \citep{wood02}, making it an efficient disc diagnostic \citep{oliveira04}. 

\begin{figure*}
\includegraphics[height=7.5cm]{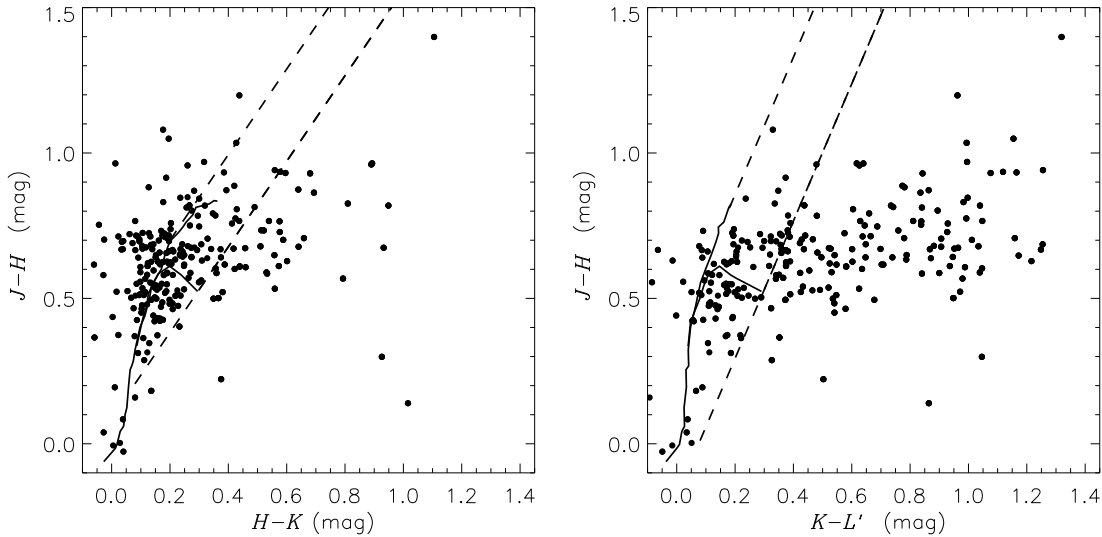}
\caption{Intrinsic colour-colour diagrams for PMS objects in NGC\,6611: $JHK$ (left) and $JHKL'$ (right). The error bars are at most 0.15\,mag in $(K-L')$ and less than 0.05\,mag for the other colours. Full lines are the loci of main-sequence and giant stars while the dashed lines are the reddening band \citep{bessell88}. Objects to the right of the reddening band for spectral type M3 have an excess indicative of the presence of a circumstellar disc. The spread in $(J-H)$ colours is consistent with the spread in visual extinction observed in the area \citep{dewinter97}.}
\label{ccd}
\end{figure*} 

Fig.\,\ref{ccd} shows the $JHK$ and $JHKL'$ colour-colour diagrams for the sample of PMS stars in NGC\,6611. The full lines are the main-sequence and giant loci; the dashed lines are the reddening band \citep{bessell88}. We assume that the spectral type of our lowest-mass object is M3 (to place the reddening band). We corrected the observed PMS colours for the cluster's average extinction. Objects to the right of the reddening band cannot have purely photospheric colours and are identified as having a $K$ and/or $L'$ excess, and therefore a circumstellar disc. Simply by counting such objects in Fig.\,\ref{ccd} we see that 45 out of 259 objects (17\%) have an ($H-K$) excess and 135 out of 259 objects (52\%) have a ($K-L'$) excess. 

Correcting for 5\% contamination from field stars (that would not show an excess), we estimate that 18\%$\pm$3\% and 55\%$\pm$5\% of the PMS objects in NGC\,6611 have ($H-K$) and ($K-L'$) excesses respectively. If we divide the PMS objects in two mass bins, 0.45\,M$_{\odot}$\,$<$\,M\,$<$\,2\,M$_{\odot}$ and 2\,M$_{\odot}$\,$<$\,M\,$<$\,7\,M$_{\odot}$, the ($K-L'$) disc frequencies are respectively 58\%$\pm$5\% and 30\%$\pm$10\%, i.e. the disc frequency is lower for more massive stars as it is observed in other star forming regions \citep*{haisch01a}. We are being conservative in our choice of the latest spectral type in the sample. Furthermore, this method tends to underestimate the fraction of early spectral type objects with excesses. Thus, our derived ($K-L'$) disc fraction is a lower limit to the true disc frequency for solar-mass stars in NGC\,6611.

\begin{table}
\centering
\caption{Comparison of the disc frequency of NGC\,6611 with similarly aged clusters. For each cluster the age, spectral type of the most massive star, $(K-L')$ disc frequency, mass range and number of stars for the determinations are given.}
\begin{tabular}{lcclrc}
\hline
cluster  &age  &massive& \hspace*{-3mm} ($K-L'$) disc          &mass             &N\\
         &Myr  &star   &frequency     &M$_{\odot}$&stars\\
\hline
IC\,348  &2.3  & B5    &65\%$\pm$8\% & 0.2$-$4          &\llap{1}07\\
NGC\,6611&2$-$3& O4    &30\%$\pm$10\%& 2$-$7            &30\\
&&                     &58\%$\pm$5\% & \llap{0}.45$-$2         &\llap{2}29\\
NGC\,2264&3.2  & O7    &52\%$\pm$10\%& $>$0.85          &56\\
\hline
\end{tabular}
\label{t1}
\end{table}

In Table\,\ref{t1} we compare the derived disc frequency with published data for nearby star forming regions of a similar age, but where the PMS stars are subject to less ionising radiation from the massive stars. Disc frequencies for IC\,348 and NGC\,2264 are respectively from \citet{haisch01a} and \citet{haisch01b}. We should point out that these disc frequencies were determined in the same way as in this letter. Without taking into account the uncertainties in determining young cluster ages, the disc frequency for NGC\,6611 is perfectly bracketed by the disc frequencies of these two clusters. Furthermore, the ($H-K$) disc fraction in NGC\,6611 is also consistent with the ($H-K$) disc fraction in IC\,348 (21\%$\pm$5\%, \citealt{haisch01a}), indicating that the fractions of {\em accretion discs} might also be similar.

\section{Conclusions}

NGC\,6611 is a young, massive cluster that ionises the Eagle Nebula. Using optical photometry we have discovered a rich low-mass, pre-main-sequence population in the cluster. Using $JHKL'$ photometry we investigate the cluster disc frequency for solar-mass objects (0.45\,M$_{\odot}$\,$<$\,M\,$<$\,2\,M$_{\odot}$). The contamination corrected disc frequency is 58\%$\pm$5\% determined from ($K-L'$) excesses. Not only is the ($K-L'$) disc frequency perfectly comparable with the disc frequencies in IC\,348 and NGC\,2264 (two similarly aged but quieter star forming regions), but also the accretion disc frequencies are similar. This seems to indicate that, despite of the presence of more (and more luminous) O-stars in NGC\,6611, disc dissipation progresses at a rate indistinguishable from other environments. 
We cannot say that the ionising flux from the O-stars has no effect on the evolution of circumstellar discs, as an L$'$-band excess indicates the presence of a circumstellar dust disc but it does not probe the outer, cooler disc regions and thus it is practically insensitive to total disc mass \citep{wood02}. Neither does it reflect other important disc evolution processes like dust grain agglomeration and annealing. Still we find no evidence that the harsher environment of NGC\,6611 significantly {\em hastens} the dissipation of circumstellar discs around solar-mass stars, when averaged over an entire cluster.

\section*{Acknowledgments}
We thank the staff of the United Kingdom Infrared Telescope (UKIRT) and Isaac Newton Telescope (INT) for their support. The UKIRT is operated by the Joint Astronomy Centre on behalf of the UK Particle Physics and Astronomy Research Council (PPARC). This publication makes use of data products from the Two Micron All Sky Survey, which is a joint project of the University of Massachusetts and the Infrared Processing and Analysis Center/California Institute of Technology, funded by the National Aeronautics and Space Administration and the National Science Foundation. Data reduction was performed using computer facilities at Keele University funded by PPARC. JMO and SPL acknowledge financial support from PPARC.

\bsp

\label{lastpage}

\end{document}